\documentclass[pra,twocolumn,showpacs,floatfix]{revtex4}
\usepackage{graphicx}
\begin{document}

\title{The absence of fragmentation in Bose-Einstein condensates}
\author{A. D. Jackson$^1$, G. M. Kavoulakis$^2$, and M. Magiropoulos$^2$}
\affiliation{$^1$The Niels Bohr International Academy, The Niels
Bohr Institute,
Blegdamsvej 17, DK-2100, Copenhagen \O, Denmark \\
$^2$Technological Education Institute of Crete, P.O. Box 1939,
GR-71004, Heraklion, Greece}
\date{\today}

\begin{abstract}

A Bose-Einstein condensate produced by a Hamiltonian which is
rotationally or translationally symmetric is fragmented as a
direct result of these symmetries. A corresponding mean-field
unfragmented state, with an identical energy to leading order
in the number of particles, can generally be constructed. 
As a consequence, vanishingly weak symmetry-breaking perturbations 
destabilize the fragmented state, which would thus be extremely 
difficult to realize experimentally, and lead to an unfragmented
condensate.

\end{abstract}
 \maketitle

{\it Introduction.} One of the most fundamental issues in the
problem of Bose-Einstein condensation is that of
fragmentation. Years after the prediction of this phase
transition by Bose and Einstein, Penrose and Onsager
\cite{Penrose} provided a rigorous criterion for the existence
of a Bose-Einstein condensate.  Starting from the $N$-particle
wave function of a system, one determines the eigenvalues of
the (Hermitian) one-body density matrix,
\begin{eqnarray}
\rho ({\bf r}_1' , {\bf r}_1 ) = \phantom{XXXXXXXXXXXXXXXXXXX}
\nonumber \\
N \, \int \, d{\bf r}_2 \ldots d{\bf r}_N \Psi^* ({\bf r}_1',
{\bf r}_2, \dots {\bf r}_N) \Psi ({\bf r}_1, {\bf r}_2, \dots,
{\bf r}_N).
\end{eqnarray}
If at least one eigenvalue is of order $N$, the system is
Bose-Einstein condensed, otherwise it is not. A single
eigenvalue of order $N$ indicates simple condensation; when
more than one of the eigenvalues are of order $N$, the
condensate is said to be fragmented \cite{Leggett}.

Years after this definition was introduced, Nozi\'eres and
Saint James \cite{Nozieres} argued that, in Hartree-Fock
approximation, the Fock term makes it energetically favorable
for the system to fragment if the effective interaction between
the bosons is attractive. The question of condensate
fragmentation remained academic, since homogeneous systems with
an effective attractive interaction are unstable against
collapse. Modern techniques for dealing with trapped cold
atoms have rekindled interest in this question since these
gases can be metastable if the effective interaction is
attractive, and it may thus be possible to realize a fragmented
state. This issue has been the subject of a number of studies
over the last decade, see e.g.,
Refs.\,\cite{WGS,Rokhsar,UL,Chris,PP,Ueda,MHUB,German,Jason,HH}.

In the present study we consider the problem of fragmentation
for an effective attractive interaction between the atoms. As
two characteristic examples we consider bosonic atoms at zero
temperature that are confined in toroidal and in harmonic
traps. We will argue that the issue of fragmentation is subtle.
For any fragmented state, it is always possible to construct a
mean-field, non-fragmented product state that has the same
energy to leading and subleading order in $N$ as the fragmented
state. In these representative problems and in any case where
the Hamiltonian is rotationally or translationally invariant,
the single-particle density matrix is diagonal, and its
eigenvalues are the occupation numbers of the corresponding
single-particle states. The system is fragmented merely as a
consequence of the symmetry of the Hamiltonian.  Due to the low
excitation energies of states of well-defined (angular)
momentum [i.e., characteristically ${\cal O}(1/N)$], such
states are fragile and virtually impossible to realize in
practice. We shall show that small symmetry-breaking terms in
the Hamiltonian that vanish in the limit $N \to \infty$ can
materially alter wave functions and can reconstruct an
unfragmented condensate.

{\it Bosons in a toroidal trap with an effective attractive
interaction.}  Consider a tight toroidal trap where the
interaction energy between the atoms is much smaller than the
excitation energy in the transverse direction.  The Hamiltonian
of this system reduces to \cite{Ueda2,GMK}
\begin{eqnarray}
  H = \sum_{i=1}^N - \frac {\partial^2} {\partial \theta_i^2}
+ V(\theta_i) - \frac 1 2 |U| \sum_{i \neq j = 1}^N
\delta(\theta_i - \theta_j), 
\label{Ham1a}
\end{eqnarray}
where $\theta$ is the angle in cylindrical polar coordinates,
$V(\theta)$ is the potential that acts along the toroidal trap,
and $U$ describes the coupling and is proportional to the
scattering length for elastic atom-atom collisions. The
dimensionless parameter $\gamma = - (N-1) |U|/(2 \pi)$ gives
the ratio between the interaction and the kinetic energy. As
shown in Refs.\,\cite{Ueda2,GMK}, the gas becomes localized and
forms a bright solitary wave when $\gamma < \gamma_c = -1/2$.

It is convenient to expand the eigenfunctions of this
Hamiltonian in the basis of states $\exp(i q \theta)/\sqrt{2
\pi}$, with $q=0, \pm1, \pm 2 \dots$, which are eigenstates 
of the angular momentum operator, $\hat{L} = - i \partial/
\partial \theta$, with eigenvalues $q$. For clarity, we truncate 
the space and consider only single particle states with $q = 
-1,\ 0$, and 1. We also use the notation for the corresponding 
Fock states
\begin{eqnarray}
   |m \rangle = |(-1)^{m}, 0^{N - 2m - L}, (+1)^{m + L} \rangle,
\end{eqnarray}
where $m$, $N - 2m - L$ and $m+L$ are the occupancies of the
single-particle states with $q = -1, 0$, and 1, respectively.
Clearly, the above states are eigenfunctions of the angular
momentum operator $\hat L$ and of the number operator $\hat N$.

The Hamiltonian of Eq.\,(\ref{Ham1a}) commutes with $\hat N$.
If $V(\theta)$ is zero or constant, it also commutes with $\hat
L$. For a specific $N$ and $L$, the eigenstates $|\psi_L \rangle$ 
of this Hamiltonian can be expanded in the basis of the states $|m
\rangle$,
\begin{eqnarray}
   |\psi_{L} \rangle = \sum_m d_m^{L} 
   |(-1)^{m}, 0^{N - 2m - L}, (+1)^{m + L} \rangle.
\label{eigequ}
\end{eqnarray}
The eigenvalue equation $H |\psi_{L}\rangle = E(L) |\psi_L \rangle$, 
where $E(L)$ is the eigenenergy, has the form \cite{JKMR}
\begin{eqnarray}
   H_{m,m} d_m^{L} + H_{m,m-1} d_{m-1}^{L} + H_{m,m+1}
   d_{m+1}^{L} = E(L) d_m^{L},
\label{eigeqq}
\end{eqnarray}
where $H_{n,m} = \langle n | H | m \rangle$ are the matrix
elements of the Hamiltonian between the states $|n\rangle$ and
$|m\rangle$. In the specific truncated space, only the matrix
elements $H_{m,m}$ and $H_{m,m \pm 1}$ are nonzero. In the 
limit $N \to \infty$, numerical solution to this equation
yields 
\begin{eqnarray}
  E(L) = - \frac {9 N} {14} + 0.488 + \frac 3 2 \frac {L^2} N 
  + {\cal O}(1/N).
\label{ennum}
\end{eqnarray}

For large $N$, it is convenient to regard $m$ as a continuous
variable. Assuming that $d_m^L$ is a differentiable function of
$m$ up to any order,
\begin{eqnarray}
   d_{m \pm 1}^L \approx d_m^L \pm \partial_m d_m^L
+ (1/2) \partial_m^2 d_m^L,
\end{eqnarray}
and the matrix eigenvalue equation can be written as a familiar
harmonic oscillator problem,
\begin{eqnarray}
  - \frac 1 {2 \mu} \partial_m^2 d_m^L
 + \left[ E_0^L + \frac \lambda 2 (m - m_0^L)^2 \right] d_m^L
= E(L) d_m^L.
\label{diffeq}
\end{eqnarray}
The solution of this equation, with the boundary condition
that $d_m^L$ vanishes as $m \to \pm \infty$, has the expected 
Gaussian form
\begin{eqnarray}
  d_m^L \propto \exp{[- \sqrt{\lambda \mu} (m - m_0^L)^2/2]}.
\label{coeff}
\end{eqnarray}
For the specific value of $\gamma = -1$ we find that $m_0^L 
= 0.1429N - L/2$, $\lambda = 14/N$, and $1/\mu = 0.2057 N$. 
The root mean square deviation of $m$ from $m_0^L$ scales like 
$\sqrt N$. The number of states participating actively in
Eq.\,(\ref{eigequ}) is ${\cal O}( \sqrt{N})$, and the 
approximation of Eq.\,(\ref{diffeq}) is justified in the 
large $N$ limit. Similarly, we find $E_0^L \approx -9N/14
-0.357 +(3/2)L^2/N$. Our approximation to the energy, $E(L) 
= E_0^L + (1/2) \sqrt{\lambda/\mu}$, thus agrees with the
exact result of Eq.\,(\ref{ennum}) for all terms shown. 
The error in energies obtained with this smooth approximation 
is ${\cal O}(1/N)$ for all $|L| < {\cal O}(\sqrt N)$ to be
considered below.
%

The one-body density matrix corresponding to the state $|\psi_L 
\rangle$ described by Eqs.\,(\ref{eigequ}) and (\ref{coeff}) is 
diagonal in the basis of angular momentum eigenstates. This is 
a direct consequence of the fact that the states $|\psi_L \rangle$ 
are eigenfunctions of the total angular momentum.  The diagonal
elements are the occupancies of the three single-particle states 
with $q = -1, 0$ and 1. Since the average value of $m = m_0$
is of order $N$ (but not equal to $N$), all three eigenvalues
are of order $N$.  The condensate is fragmented.

We now consider the effect of a very weak inhomogeneous
potential, $V(\theta)$, on the eigenvalues of the
single-particle density matrix.  (This point has also been
addressed in Ref.\,\cite{Ueda}.) As we will see, the state is
no longer fragmented in the presence of such an inhomogeneity.
We choose $V(\theta) \propto \delta \cos \theta$, with $\delta
\ll 1$. This potential connects single particle states with
$\Delta q = \pm 1$,
\begin{eqnarray}
  V(\theta) = \delta (a_0 a_1^{\dagger} + a_1 a_0^{\dagger}
 + a_0 a_{-1}^{\dagger} + a_{-1} a_0^{\dagger}),
\end{eqnarray}
with $a_q$ and $a_q^{\dagger}$ the usual annihilation and
creation operators of particles with angular momentum $q$.  
For sufficiently weak $V( \theta )$, the resulting eigenstates 
$|\psi \rangle$ of the Hamiltonian, Eq.\,(\ref{Ham1a}), 
can be expressed as a linear superposition of the states 
$|\psi_L \rangle$,
\begin{eqnarray}
   |\psi \rangle = \sum_{m,L} d_m^L |(-1)^{m}, 0^{N - 2m - L},
   (+1)^{m + L} \rangle.
\end{eqnarray}
Given the form of $V(\theta)$, it can connect the state
$|(-1)^m, 0^{N-2m-L}, (+1)^{m+L} \rangle$ to the following four
states
\begin{eqnarray}
|1\rangle &=&|(-1)^{m-1}, 0^{N-2m-L+1}, (+1)^{m+L} \rangle,
\nonumber \\
|2\rangle &=&|(-1)^{m+1}, 0^{N-2m-L-1}, (+1)^{m+L} \rangle,
\nonumber \\
|3\rangle &=&|(-1)^m, 0^{N-2m-L-1}, (+1)^{m+L+1} \rangle,
\nonumber \\
|4\rangle &=&|(-1)^m, 0^{N-2m-L+1}, (+1)^{m+L-1} \rangle.
\end{eqnarray}
Thus, Eq.\,(\ref{eigeqq}) assumes the form,
\begin{eqnarray}
H_{m,m} d_m^L + H_{m,m-1} d_{m-1}^L &+& H_{m,m+1} d_{m+1}^L +
\nonumber \\
+ \delta \sqrt{{\overline m}_0 (N - 2 {\overline m}_0)}
( d_m^{L+1} + d_m^{L-1} &+& d_{m+1}^{L-1} + d_{m-1}^{L+1}) 
\nonumber \\
&=& E d_m^L.
\end{eqnarray}
Here, $\sqrt{{\overline m}_0 (N - 2 {\overline m}_0)}$ approximates
the value of this prefactor with its value at the minimum,
i.e., ${\overline m}_0 = m_0^{L=0} = 0.143153 N$.

Converting this matrix eigenvalue equation into a differential
equation as above, we find
\begin{eqnarray}
 &-&  \frac 1 {2 \mu} \partial_m^2 d_m^L
 + \left[ E_0^L + \frac \lambda 2 (m - m_0^L)^2 \right] d_m^L +
\nonumber \\
  &+&  \delta \sqrt{{\overline m}_0 (N - 2 {\overline m}_0)}
 (4 d_m^L + 2 \partial_{LL} d_m^L +
\nonumber \\
 &+&  \partial_{mm} d_m^L - 2 \partial_L \partial_m d_m^L )
 = E d_m^L.
\end{eqnarray}
The solution of this differential equation has the form
\begin{equation}
d_m^L \propto \exp{[-a_1(m + L/2 - {\overline m}_0)^2-a_2 L^2]},
\end{equation}
with $a_1$ and $a_2$ positive, since this function
vanishes as $m$ and $L$ tend to infinity in any direction.
Numerical calculations with a symmetry-breaking term
$V(\theta)$ of the form $\delta = - (1/100) (100/N)^{1.15}$
verify that the coefficients $d_m^L$ are indeed Gaussian distributed
as a function of $m$ and $L$.

Direct calculation reveals that the one-body density matrix now
has only one eigenvalue of order $N$. [The next-largest
eigenvalue is ${\cal O}(N^{0.576})$.] This result is readily
understood as a consequence of the Gaussian support of the
$d_m^L$. If the symmetry-breaking potential, $V(\theta)$, is
sufficiently strong that the root mean square variations in $m$
and $L$ grow with $N$ but small enough that $\Delta m/\langle m
\rangle$ and $\Delta L/ \langle L \rangle$ vanish as $N \to
\infty$, the matrix elements of the one-body density matrix,
$\rho_{ij} = \langle a^{\dag}_i a_j \rangle$, are $\sqrt{n_i
n_j}$, with $n_i$ the occupation number of single particle
state $i$. The one-body density matrix is thus rank one
separable. It has one non-zero eigenvalue of $\sum_{j} \, n_j =
N$; the elements of the corresponding eigenvector are 
proportional to $\sqrt{n_j}$.  All other eigenvalues are zero, 
and the condensate is unfragmented.

The symmetry-breaking potential, $V(\theta)$, must be
sufficiently strong if it is to yield the desired mixing of the
states $|\psi_L \rangle$. According to Eq.\,(\ref{ennum}), the 
states with $L \ne 0$ are separated from the $L=0$ ground state 
by a term that scales as $L^2/N$. Since the contribution of
$V(\theta)$ to the energy is of order $N \delta$, $\delta$ must
vanish less rapidly than $1/N^2$.  In addition, low-lying
excited states (for each $L$) are separated by an energy of
${\cal O}(N^0)$ from the lowest energy state. Validity of the
truncation to the states $|\psi_L \rangle$ requires that $\delta$
vanishes more rapidly than $1/N$.  Clearly, the second
condition is dictated by approximations made in this
calculation and not with the absence of fragmentation.  In
short, even symmetry-breaking potentials which vanish in the
large $N$ limit are sufficient to ensure that the condensate is
not fragmented.

{\it Bosons in a harmonic trap with an effective attractive
interaction.}  We now turn to two additional systems which have
been considered as examples of condensate fragmentation. First,
consider rotating bosonic atoms confined in a two-dimensional
harmonic trap and subject to the Hamiltonian \cite{WGS}. In 
cylindrical polar coordinates,
\begin{eqnarray}
 H =  \sum_{i=1}^N - \frac 1 2 \nabla_i^2
  + \frac {1} 2 \rho_i^2
  - \frac 1 2 |{\eta}| \sum_{i \neq j = 1}^N \delta({\bf r}_i - {\bf r}_j).
\label{Ham1}
\end{eqnarray}
Here $\eta$, which describes the atom-atom interaction, is 
proportional to
the s-wave scattering length.  If the coupling is weak, $|\eta|
\ll 1$, this Hamiltonian can be truncated to include only
states in the lowest Landau level with zero radial nodes and
$m$ quanta of angular momentum, $\phi_{m} = z^m e^{-|z|^2/2}/
\sqrt{\pi m!}$, with $z = x + i y$, where $x$ and $y$ are the
Cartesian coordinates.  In this case, as shown by
Wilkin, Gunn and Smith \cite{WGS} and by Mottelson \cite{Ben},
the interaction energy of the lowest energy state for any given
angular momentum is the same as that of the non-rotating
system,
\begin{eqnarray}
E(L) = - \frac {|\eta|} 2 N (N-1) \int |\phi_{0}|^4  d^2
\rho =  - \frac {|\eta|} {4 \pi} N (N-1).
\label{energyexact}
\end{eqnarray}
The full energy of these states contains an additional
contribution of $|L|+1$ from the confining potential.  The
corresponding exact many-body eigenstate describes a center of
mass excitation with
\begin{eqnarray}
  \Psi_{\rm ex}^L({z}_1, {z}_2, \dots, {z}_N) = {\cal N}_L {Z}^L
  \sum_{i=1}^N \exp(-|{z}_i|^2/2) .
\end{eqnarray}
Here, ${\cal N}_L = 1/\sqrt{\pi^N N^L L!}$ and $Z$ is
the center of mass coordinate, i.e., ${Z} = \sum_{i=1}^N {z}_i$.

The eigenvalues of the single-particle density matrix are
$\rho_m = (N-1)^{L-m} L!/[N^{L-1} (L-m)! m!]$ \cite{WGS}.
Due to the axial symmetry of the Hamiltonian, this density 
matrix is diagonal, and its eigenvalues are simply the
occupation numbers, $N |c_m|^2$, of the single-particle states.
The energy is minimized when all $c_m$ have the same phase,
which can be taken as positive without loss of generality. In
the limit of infinite $N$ and $L$ with $l = L/N$
finite, we see that \cite{Benp} $|c_m|^2 (l) = {l^m}
\exp(-l)/{m!}$. According to the usual criterion, this is a
fragmented state.

It is possible, however, to construct a mean-field, product
wave function which has the same interaction energy and which
is necessarily unfragmented.  Consider the simple form
\begin{eqnarray}
  \Psi_{\rm MF}^l(z_1, z_2, \dots, z_N) =
  \prod_{i=1}^N \sum_{m=0}^{\infty} c_m \phi_{m}(z_i),
\end{eqnarray}
with the coefficients $c_m = \sqrt{l^m/m!} \exp(-l/2)$.  This
state is normalized, and the expectation value of the angular
momentum per particle is $l = L/N$. The interaction energy of
this state can be calculated analytically,
\begin{eqnarray}
E_{\rm MF}^l = -\frac {|\eta|} 2 N (N-1)
  \int |\sum_{m=0}^{\infty} c_m \phi_{m}(z)|^4 dx \, dy
\nonumber \\
= - \frac {|\eta|} {4 \pi} N (N-1) e^{- 2 l}
  \sum_{m=0}^{\infty} \frac 1 {m!} \left( \frac l 2 \right)^m
  \sum_{k, j=0}^m
  \left( \begin{array}{c} m \\ k \end{array} \right)
  \left( \begin{array}{c} m \\ j \end{array} \right)
\nonumber \\
=  - \frac {|\eta|} {4 \pi} N (N-1)  e^{- 2 l}
  \sum_{m=0}^{\infty} \frac 1 {m!} \left( \frac l 2 \right)^m
  4^m
\nonumber \\
= - \frac {|\eta|} {4 \pi} N (N - 1),
\label{man}
\phantom{XXXXXXXXXXXXXXXXX}
\end{eqnarray}
which is identical to the energy given by
Eq.\,(\ref{energyexact}). The overlap between the states
$\Psi_{\rm ex}^L$ and $\Psi_{\rm MF}^l$ can also be calculated
analytically as
\begin{eqnarray}
 \langle \Psi_{\rm ex}^L | \Psi_{\rm MF}^l \rangle =
 \left( \frac L {N e} \right)^{L/2} \frac {\pi^{N/2} N^L}
 {\sqrt{\pi^N N^L L!}}
\approx \frac 1 {\sqrt {2 \pi L}}.
\end{eqnarray}
This overlap vanishes in the thermodynamic limit $L \to
\infty$. This comes as no surprise, since $\Psi_{\rm ex}^L$
describes a state that is spread uniformly around the center 
of the trap, while $\Psi_{\rm MF}^l$ describes precisely the 
non-rotating ``clump'' of matter displaced from the center 
of the trap and rotating around it.

The existence of a mean field state with an energy close to the
exact eigenvalue is a relatively general consequence of
rotational or translational invariance.  Given a Hamiltonian
which is axially or translationally invariant, the one-body
density matrix is diagonal with eigenvalues equal to the
occupation numbers of the corresponding single particle states.
From these occupancies, it is possible to construct a
mean-field wave function with the same energy as the exact
solution to leading and often subleading order in the number of
particles, $N$. The same conclusion applies to the argument of
Nozieres and Saint James: The exchange interaction does not
necessarily favor fragmentation since there exists a
non-fragmented mean-field product state with the same energy in
the $N \to \infty$ limit.  As seen in the case of toroidal
confinement, the question of whether the system is better
described by a wave function which is an eigenfunction of the
total momentum or angular momentum (and thus fragmented) or is
better described by a mean field wave function (and thus not
fragmented) must depend on the response of the system to
vanishingly small symmetry-breaking terms.

This issue can be investigated with arguments and conclusions
identical to those above.  A one-body symmetry-breaking term,
$V( z ) = \delta ( z + z^* )$ is introduced, where again $z=x+iy$.  
The basis of states is truncated to include only the lowest-energy 
states, $\Psi_{\rm ex}^L$.  Evidently, $V(z)$ can only connect 
the state $L$ with the states $L \pm 1$. The corresponding matrix 
elements are, e.g.,
\begin{equation}
\langle  \Psi_{\rm ex}^{L+1} | V |  \Psi_{\rm ex}^L \rangle =
\delta \sqrt{N (L+1)} \ \ {\rm for}\ \ L \ge 0.
\label{delV}
\end{equation}
If $\delta$ vanishes with increasing $N$, this truncation of
states is legitimate.  If it vanishes more slowly than
$1/\sqrt{N}$, there will be significant mixing of the states
$\Psi_{\rm ex}^L$. As in the case of toroidal confinement, the
wave function will have localized (i.e., Gaussian) support in
the space of single particle states. Precisely as before, only
one eigenvalue of the one-body density matrix is of order $N$,
and the condensate is not fragmented.

Identical arguments can be applied to the related but simpler
two-state model of Ref.\,\cite{Chris}. There, the condensate is
fragmented due to a ``parity'' symmetry. This is reflected in
the fact that eigenstates contain, e.g., only an even (or odd)
number of particles in one of the states. The energy difference 
between the lowest-energy even and odd states vanishes 
exponentially with $N$. Once again, a vanishingly small one-body
symmetry-breaking term, proportional to $(a^{\dag}_0 a_1 +
a^{\dag}_1 a_0)$, is sufficient to reconstruct an unfragmented
condensate. These two examples will be described in greater
detail elsewhere. Finally, similar arguments apply to the
studies of Refs.\,\cite{Jason,HH}, where a state fragmented 
by some symmetry of the Hamiltonian can be restored to a simple 
unfragmented condensate by very weak symmetry-breaking perturbations.

{\it General conclusions.} For many systems of a large but
finite number of bosons with attractive interactions, mean
field theory provides a good description of the ground state
energy and leads to the unambiguous prediction of an
unfragmented condensate. The imposition of general constraints,
such as conserved total momentum or angular momentum,
characteristically produces minimal changes in the ground state
energy and frequently indicates condensate fragmentation. This
apparent contradiction has led some authors to suggest modified
criteria for condensate fragmentation. We have offered an
alternate resolution. The various examples considered here all
suggest that the small excitation energies of excited states in
these systems can render them sensitive to vanishingly small
symmetry-breaking perturbations. The resulting localized (i.e.,
often Gaussian) support of the wave function then leads to a
one-body density matrix, approximately given as $\rho_{ij} =
\sqrt{n_i n_j}$, which is rank one separable with one
eigenvalue of ${\cal O}(N)$ in the $N \to \infty$ limit. The
unfragmented condensate, deconstructed by rigorous symmetries,
can be reconstructed by small symmetry-breaking perturbations.
Such perturbations can be difficult to eliminate
experimentally.  (Such antagonism between the mean field
approximation and symmetries is well known. For example,
insistence on maintaining translational invariance in fermion
systems leads inevitably to a trivial Hartree-Fock wave
function of plane wave states and a poor description of both
the ground state energy and wave function.) While the present
results in no sense rule out the possibility of condensate
fragmentation in systems of bosonic atoms, they do suggest that
it is important to demonstrate that theoretical indicators of
fragmentation are robust with respect to small
symmetry-breaking perturbations.


\begin{thebibliography}{99}

\bibitem{Penrose} O. Penrose and L. Onsager, Phys. Rev. {\bf 104},
576 (1956).

\bibitem{Leggett} A. J. Leggett, Rev. Mod. Phys. {\bf 73},
307 (2001).

\bibitem{Nozieres} P. Nozi\'eres and D. Saint James, J. Physique
{\bf 43}, 1133 (1982).

\bibitem{WGS} N. K. Wilkin, J. M. F. Gunn, and R. A. Smith, Phys. Rev. Lett.
{\bf 80}, 2265 (1998).

\bibitem{Rokhsar} D. Rokhsar, e-print arXiv:cond-mat/9812260.

\bibitem{UL} M. Ueda and A. J. Leggett, Phys. Rev. Lett. {\bf 83},
1489 (1999).

\bibitem{Chris} \O. Elgar\o y and C. J. Pethick, Phys. Rev. A {\bf 59},
1711 (1999).

\bibitem{PP} C. J. Pethick and L. P. Pitaevskii, Phys. Rev. A {\bf 62},
033609 (2000).

\bibitem{Ueda} R. Kanamoto, H. Saito, and M. Ueda, Phys. Rev.
Lett. {\bf 94}, 090404 (2005).

\bibitem{MHUB} E. J. Mueller, T.-L. Ho, M. Ueda, and G. Baym,
Phys. Rev. A {\bf 74}, 033612 (2006).

\bibitem{German} K. Sakmann, A. I. Streltsov, O. E. Alonz, and
L. S. Cederbaum, e-print arXiv:0802.3417.

\bibitem{Jason} T.-L. Ho and S. K. Yip, Phys. Rev. Lett. {\bf 84},
4031 (2000).

\bibitem{HH} Xia-Ji Liu, Hui Hu, Lee Chang, Weiping Zhang, Shi-Qun Li,
Yu-Zhu Wang, Phys. Rev. Lett. {\bf 87}, 030404 (2001).

\bibitem{Ueda2} R. Kanamoto, H. Saito, and M. Ueda,
Phys. Rev. A {\bf 68}, 043619 (2003).

\bibitem{GMK} G. M. Kavoulakis, Phys. Rev. A {\bf 69},
023613 (2004).

\bibitem{JKMR} A. D. Jackson, G. M. Kavoulakis, B. Mottelson,
and S. M. Reimann, Phys. Rev. Lett. {\bf 86}, 945 (2001).

\bibitem{Ben} B. Mottelson, Phys. Rev. Lett. {\bf 83}, 2695 (1999).

\bibitem{Benp} B. Mottelson, private communication.

\end{thebibliography}
\end{document}